\documentclass[prd,showpacs,preprintnumbers,amsmath,amssymb]{revtex4}
\usepackage{graphics,color,subfigure}
\usepackage{epsfig}
\usepackage{dcolumn}% Align table columns on decimal point
\usepackage{bm}% bold math

\begin{document}
%\begin{CJK}{GBK}{song}
\title{Possible $\Lambda_c\Lambda_c$ molecular bound state}
\author{Wakafumi Meguro}
\author{Yan-Rui Liu}\email{yrliu@th.phys.titech.ac.jp}
\author{Makoto Oka}\email{oka@th.phys.titech.ac.jp}
\affiliation{Department of Physics, H-27, Tokyo Institute of Technology, Meguro,
Tokyo 152-8551, Japan}

\date{\today}

\begin{abstract}

Possible $\Lambda_c\Lambda_c$ hadronic molecule is investigated in the one-pion-exchange potential model. In the study with this model, the heavier meson exchange effects are encoded into a phenomenological cutoff parameter and couplings to the nearby $\Sigma_c\Sigma_c$, $\Sigma_c\Sigma_c^*$, and $\Sigma_c^*\Sigma_c^*$ channels are essential. From the numerical results, we find that a molecular bound state of two $\Lambda_c$'s is possible, where the tensor force plays a crucial role, although the binding energies are sensitive to the cutoff parameter.

\end{abstract}


%\end{CJK}
\maketitle

%%%%%%%%%%%%%%%%%%%%%%%%%%%%%%%%%%%%%%%%%%%
%\section{Introduction}\label{sec1}
%%%%%%%%%%%%%%%%%%%%%%%%%%%%%%%%%%%%%%%%%%%

Hadronic molecules are loosely bound states of hadrons, whose inter-hadron distances are larger than the quark confinement size. The deuteron is the well-established molecule composed of a proton and a neutron. The triton, hypertriton and so on are also regarded as molecular bound states of the light baryons. Recently observed near-threshold charmonium-like mesons triggered lots of studies on the molecule problem in the heavy quark realm. In this note, we consider the molecule problem of two $\Lambda_c$'s.

Compared with light baryons, the heavy quark baryons are more likely to be bound. One reason is the larger reduced mass of the system. The relatively small kinetic term in the Hamiltonian is advantageous for the bound state. The other reason is the heavy quark spin symmetry and thus the importance of channel coupling. In the limit of infinitely heavy quark, QCD interaction manifests heavy quark flavor symmetry and heavy quark spin symmetry. The latter symmetry leads to degenerate $\Sigma_c$ and $\Sigma_c^*$. In the real world, the mass difference between $\Sigma_c$ and $\Sigma_c^*$ is indeed smaller than that between $\Sigma$ and $\Sigma^*$. Since the coupled channel effects become important when two channels are closer, it is necessary to include such effects in the study of heavy quark baryon interactions. We explore the importance of such effects in the $\Lambda_c\Lambda_c$ molecule problem.

\begin{table}[htb]
\begin{tabular}{cccccc}\hline
Channels & 1 & 2 & 3 & 4 & 5 \\\hline
$J^P=0^+$&$\Lambda_c\Lambda_c(^1S_0)$&$\Sigma_c\Sigma_c (^1S_0)$&$\Sigma_c^*\Sigma_c^* (^1S_0)$& $\Sigma_c^*\Sigma_c^* (^5D_0)$ & $\Sigma_c\Sigma_c^* (^5D_0)$\\
\hline
\end{tabular}
\caption{The $S$-wave $\Lambda_c\Lambda_c$ state and the channels which couple to it.}\label{chns}
\end{table}

The quantum numbers of the $S$-wave $\Lambda_c\Lambda_c$ are $I(J^P)=0(0^+)$. Here we consider in total five channels which are given in Table \ref{chns}. The wave function of the 5-th channel is taken to be
%\begin{eqnarray}\label{wave5}
%|\Sigma_c\Sigma_c^*\rangle=\frac{1}{\sqrt6}[(\Sigma_c^{++}\Sigma_c^{*0}-\Sigma_c^+\Sigma_c^{*+}+\Sigma_c^0\Sigma_c^{*++})-
%(\Sigma_c^{*0}\Sigma_c^{++}-\Sigma_c^{*+}\Sigma_c^++\Sigma_c^{*++}\Sigma_c^0)],
%\end{eqnarray}
\begin{eqnarray}\label{wave5}
|\Sigma_c\Sigma_c^*\rangle=\frac{1}{\sqrt2}\Big([\Sigma_c\Sigma_c^*]^{I=0}_{S=0}-
[\Sigma_c^*\Sigma_c]^{I=0}_{S=0}\Big),
\end{eqnarray}
where the minus sign comes from the exchange of two Fermions.

In Ref. \cite{Liu2011}, we have explored $\Lambda_cN$ system by using both one-pion-exchange potential (OPEP) model and one-boson-exchange potential model where exchanges of scalar and vector mesons are also included. It is observed that the two models may give consistent binding energies and corresponding radii. In the OPEP model, the contributions from shorter distance interactions are encoded in a phenomenological cutoff parameter. Since there is no pion exchange in the $\Lambda_c\Lambda_c$ channel, the possible binding solution must result from the coupled channel effects. In the present study, we use OPEP model and investigate whether long range interaction may lead to a molecular bound $\Lambda_c\Lambda_c$ state.

Besides the five channels in Table \ref{chns}, $\Xi_{cc}N$ may also contribute. But its contribution may be important only at short distance since the exchanged mesons between $\Xi_{cc}N$ and any channel in Table \ref{chns} are much heavier. Here we neglect the $\Xi_{cc}N$ channel as we are considering the possibility of loosely bound molecule. It was proposed that a bound state may exist also in the $\Xi_{cc}N$ system in Ref. \cite{Riska2005}.

The interaction Lagrangian reads \cite{Yan1992}
\begin{eqnarray}
{\cal L}_{int}&=&g_1{\rm tr}(\overline{B}_6\gamma_\mu\gamma_5A^\mu B_6)+[g_2{\rm tr}(\overline{B}_6\gamma_\mu\gamma_5A^\mu B_{\bar{3}})+h.c.]+[g_3{\rm tr}(\overline{B}^*_{6\mu}A^\mu B_6)+h.c.]\nonumber\\
&&+[g_4{\rm tr}(\overline{B}_6^{*\mu}A_\mu B_{\bar 3})+h.c.]+g_5{\rm tr}(\overline{B}_6^{*\nu}\gamma_\mu\gamma_5A^\mu B_{6\nu}^*)+g_6{\rm tr}(\overline{B}_{\bar 3}\gamma_\mu\gamma_5A^\mu B_{\bar 3}),
\end{eqnarray}
where
\begin{eqnarray}
&B_{\bar{3}}=\left(\begin{array}{ccc}
0& \Lambda_c^+ & \Xi_c^{+}\\
-\Lambda_c^+ & 0 & \Xi_c^{0}\\
-\Xi_c^{+}& -\Xi_c^{0} & 0 \end{array}\right), \quad
B_6=\left(\begin{array}{ccc}
\Sigma_c^{++}& \frac{1}{\sqrt2}\Sigma_c^+ & \frac{1}{\sqrt2}\Xi_c^{\prime+}\\
\frac{1}{\sqrt2}\Sigma_c^+ & \Sigma_c^{0} & \frac{1}{\sqrt2}\Xi_c^{\prime0}\\
\frac{1}{\sqrt2}\Xi_c^{\prime+}& \frac{1}{\sqrt2}\Xi_c^{\prime0} & \Omega_c^0 \end{array}\right),\quad
\Pi={\sqrt2}\left(
\begin{array}{ccc}
\frac{\pi^0}{\sqrt2}+\frac{\eta}{\sqrt6}&\pi^+&K^+\\
\pi^-&-\frac{\pi^0}{\sqrt2}+\frac{\eta}{\sqrt6}&K^0\\
K^-&\bar{K}^0&-\frac{2}{\sqrt6}\eta
\end{array}\right),&
\end{eqnarray}
$B_6^*$ is similar to $B_6$, and $A_\mu=\frac{i}{2}[\xi^\dag(\partial_\mu\xi)+(\partial_\mu\xi)\xi^\dag]$ with $\xi=\exp[\frac{i\Pi}{2f}]$ is the axial vector current. The decay constant in the chiral limit has the value $f=92.3$ MeV. In the heavy quark limit, the heavy quark spin symmetry requires a few relations for the coupling constants, i.e., $g_3=\frac{\sqrt3}{2}g_1$, $g_5=-\frac32g_1$,  $g_4=-\sqrt{3}g_2$, $g_6=0$. If one further uses the quark model symmetry, one has $g_1=-\sqrt{\frac83}\,\,g_2$. The relative phase between $g_1$ and $g_2$ is actually irrelevant \cite{Liu2011}. From the decay widths of $\Sigma_c$ and $\Sigma_c^*$ \cite{PDG2010}, we obtain $|g_2|=0.598$ and $|g_4|=0.999$ after averaging over the different charge states. In the numerical evaluation, we consistently use the following values and phases
\begin{eqnarray}
&g_2=-0.598,\quad g_4=0.999,\quad g_1=\frac{\sqrt8}{3}g_4,\quad g_3=\sqrt{\frac23}g_4,\quad g_5=-\sqrt{2}g_4.&
\end{eqnarray}

With the above Lagrangian, one derives the non-relativistic potentials. To incorporate the extended structure of baryons, a monopole type form factor $F(q)=\frac{\Lambda_\pi^2-m_\pi^2}{\Lambda_\pi^2-q^2}$ is introduced phenomenologically at each interaction vertex where $q$ is the pion 4-momentum. In principle, the cutoffs at the vertices $\Lambda_c\Sigma_c\pi$, $\Lambda_c\Sigma_c^*\pi$, $\Sigma_c\Sigma_c\pi$, $\Sigma_c\Sigma_c^*\pi$, and $\Sigma_c^*\Sigma_c^*\pi$ are different. To reduce the number of parameters and simplify the calculation, we use the approximation that these five cutoffs are equal and we label this common cutoff $\Lambda_\pi$. $\Lambda_\pi$ is poorly known but its value, around 1 GeV, may be comparable to the nuclear models \cite{Bonn,Julich}. This parameter also plays a role of compensation for the short and inter-mediate range interactions. In our study, we treat it as a free parameter and discuss whether the $\Lambda_c\Lambda_c$ molecule-like bound state is possible within the reasonable domain of $\Lambda_\pi$. We may denote the potential in the following form,
\begin{eqnarray}\label{pot-diag}
V_{ij}(\Lambda_\pi,r)=C(i,j)\frac{m_\pi^3}{24\pi f_\pi^2}\left[\vec{\cal O}_1\cdot\vec{\cal O}_2Y_1(r)+{\cal O}_{ten}H_3(r)\right],
\end{eqnarray}
where $i,\, j$ from 1 to 5 are the labels of the channels, $C(i,j)$ is the coefficient containing the coupling constants, $\vec{\cal O}_{1}$ ($\vec{\cal O}_{2}$) is the Pauli matrix $\vec{\sigma}$, the transition spin matrix $\vec{S}_t$ (or its Hermitian conjugation) explained below, or the matrix $\vec{\sigma}_{rs}\equiv-S_{t\mu}^\dag\vec{\sigma}S_t^\mu$, ${\cal O}_{ten}=\frac{3(\vec{\cal O}_1\cdot\vec{r})(\vec{\cal O}_2\cdot\vec{r})}{r^2}-(\vec{\cal O}_1\cdot\vec{\cal O}_2)$ is the tensor operator, and $Y_1$, $H_3$ and relevant functions are defined as
\begin{eqnarray}
Y(x)&=&\frac{e^{-x}}{x},\quad H(x)=(1+\frac{3}{x}+\frac{3}{x^2})Y(x),\nonumber\\
Y_1(r)&=&Y(mr)-\left(\frac{\Lambda}{m}\right)Y(\Lambda r)-\frac{\Lambda^2-m^2}{2m\Lambda}e^{-\Lambda r},\nonumber\\
H_3(r)&=&H(mr)-\left(\frac{\Lambda}{m}\right)^3H(\Lambda r)-\frac{(\Lambda^2-m^2)\Lambda}{2m^3}Y(\Lambda r)-\frac{(\Lambda^2-m^2)\Lambda}{2m^3}e^{-\Lambda r}.
\end{eqnarray}

The transition spin $S_t^\mu$ for the Rarita-Schwinger field $u^\mu$ is defined through $u^\mu=S_t^\mu \Phi$, where $\Phi$ denotes the spin wave function of $\Sigma_c^*$ defined by
\begin{eqnarray}
\Phi(3/2)=(1,0,0,0)^T,\quad \Phi(1/2)=(0,1,0,0)^T,\quad \Phi(-1/2)=(0,0,1,0)^T,\quad \Phi(-3/2)=(0,0,0,1)^T.
\end{eqnarray}
Explicitly, the time component of $S_t^\mu$ vanishes in the static limit and the other components are
\begin{eqnarray}
S_t^x=\frac{1}{\sqrt2}\left(\begin{array}{cccc}-1&0&\sqrt{\frac13}&0\\0&-\sqrt{\frac13}&0&1\end{array}\right),\quad
S_t^y=-\frac{i}{\sqrt2}\left(\begin{array}{cccc}1&0&\sqrt{\frac13}&0\\0&\sqrt{\frac13}&0&1\end{array}\right),\quad
S_t^z=\left(\begin{array}{cccc}0&\sqrt{\frac23}&0&0\\0&0&\sqrt{\frac23}&0\end{array}\right).
\end{eqnarray}

In deriving the potentials, we have neglected the $\delta$-functional term of the central force since we are considering molecule-like bound state problem. In the above potentials, $m$ ($\Lambda$) is not always $m_\pi$ ($\Lambda_\pi$). In the transition potentials $V_{15}$, $V_{25}$, $V_{35}$, and $V_{45}$, non-vanishing time component $q_0$ of the pion 4-momentum may be a better approximation. In these cases, we have $m=\sqrt{m_\pi^2-q_0^2}$ and $\Lambda=\sqrt{\Lambda_\pi^2-q_0^2}$. For the value of $|q_0|$, we use $(m_{\Sigma_c^*}-m_{\Sigma_c})/2$ for $V_{15}$ and $V_{25}$, and $(m_{\Sigma_c^*}^2-m_{\Sigma_c}^2)/(4m_{\Sigma_c^*})$ for $V_{35}$ and $V_{45}$. Note that there are two terms in the final potential $V_{55}$ due to the antisymmetrization given in Eq. (\ref{wave5}),
\begin{eqnarray}
V_{55}(\Lambda_\pi,r)=g_1g_5\frac{m^3}{24\pi f_\pi^2}\Big[Y_1(r)-2H_3(r)\Big]+|g_3|^2\frac{m^3}{24\pi f_\pi^2}\Big[  Y_1(r)+ H_3(r)\Big].
\end{eqnarray}
We use $q_0=0$ in the $g_1g_5$ part and $|q_0|=m_{\Sigma_c^*}-m_{\Sigma_c}$ in the $|g_3|^2$ part. The above $|q_0|$'s are derived in the static limit of the heavier side, initial states or final states.

\begin{figure}[htb]
\centering
\begin{tabular}{cc}
\scalebox{0.6}{\includegraphics{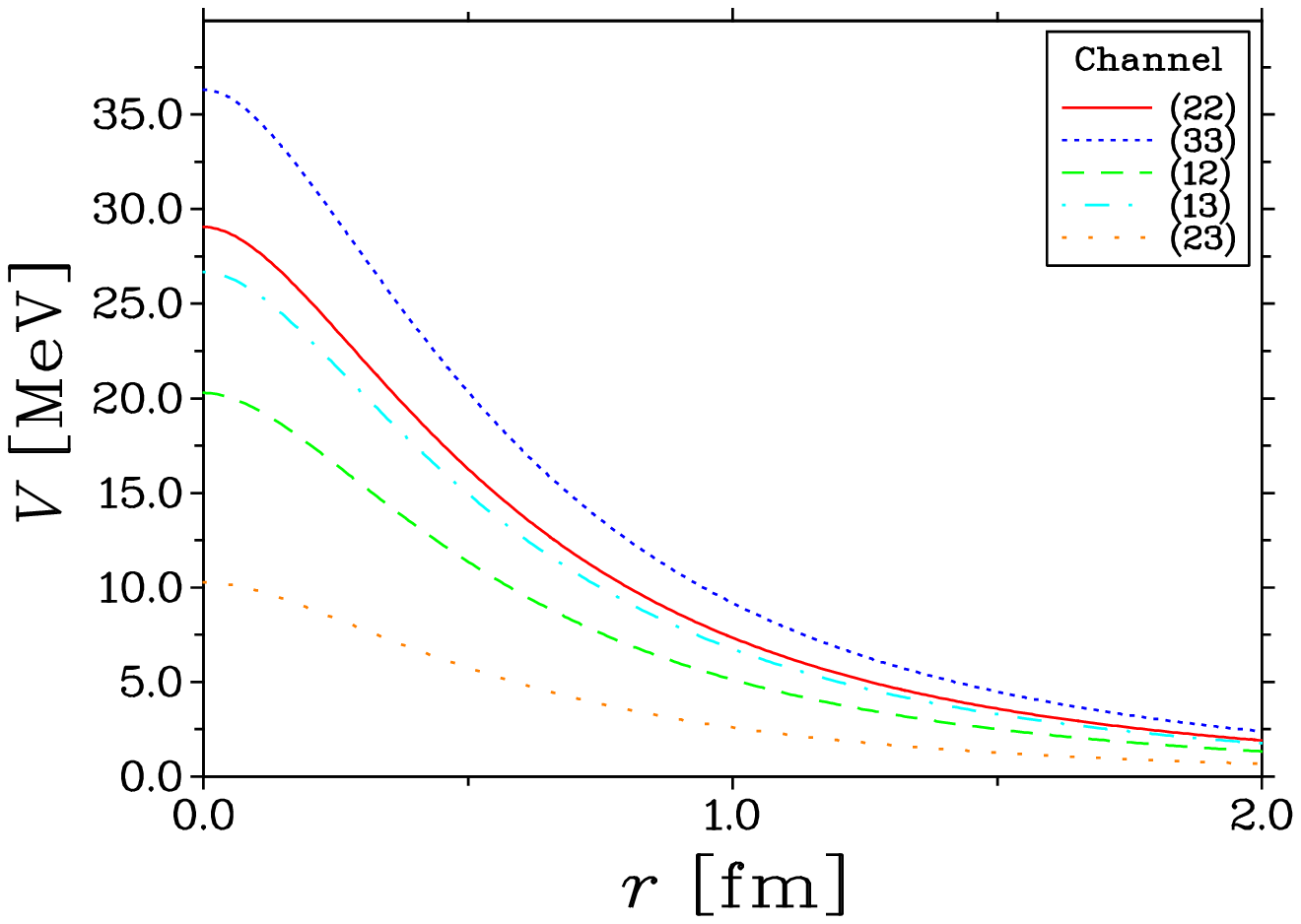}}&
\scalebox{0.6}{\includegraphics{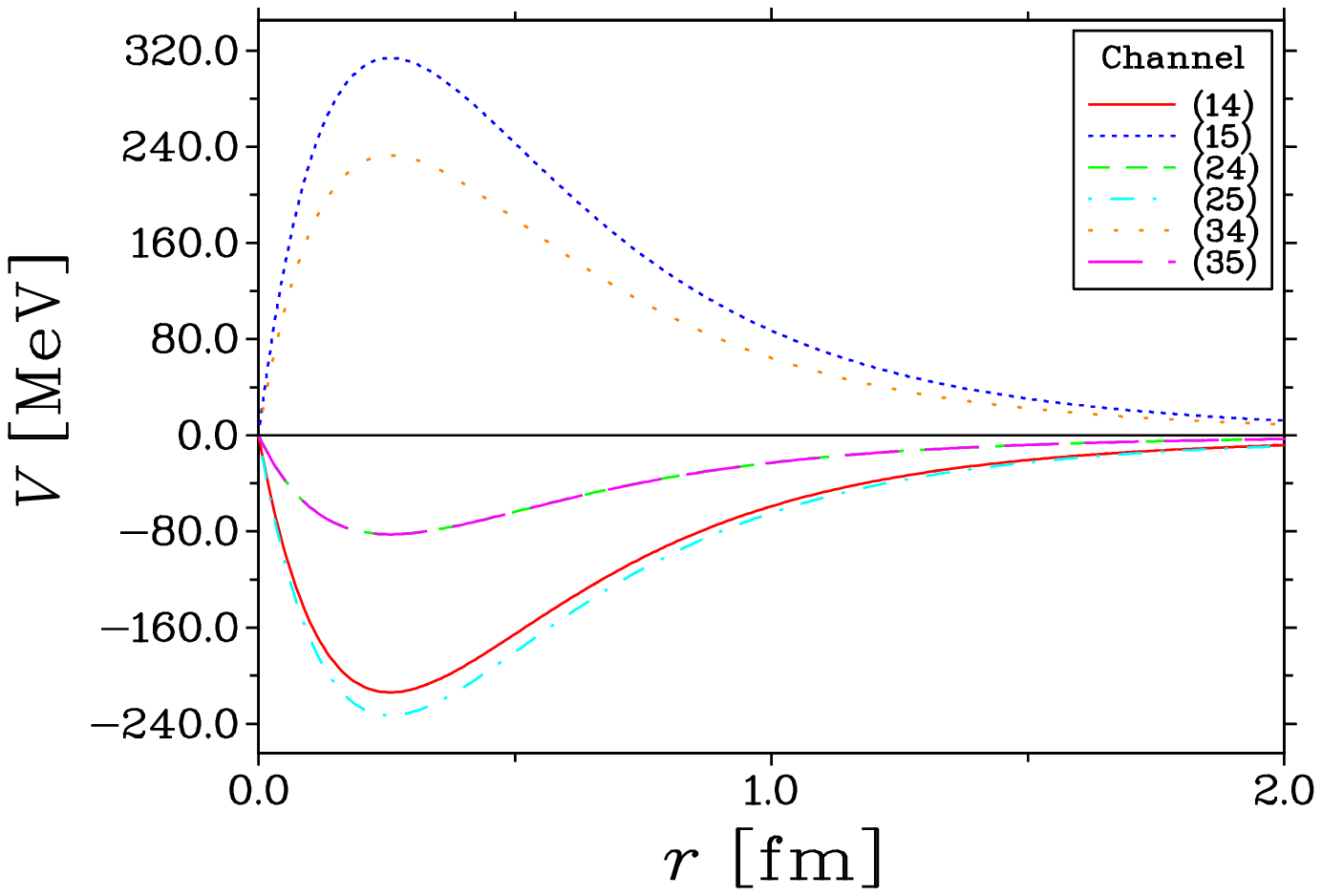}}\\
(a) & (b)\\
\scalebox{0.6}{\includegraphics{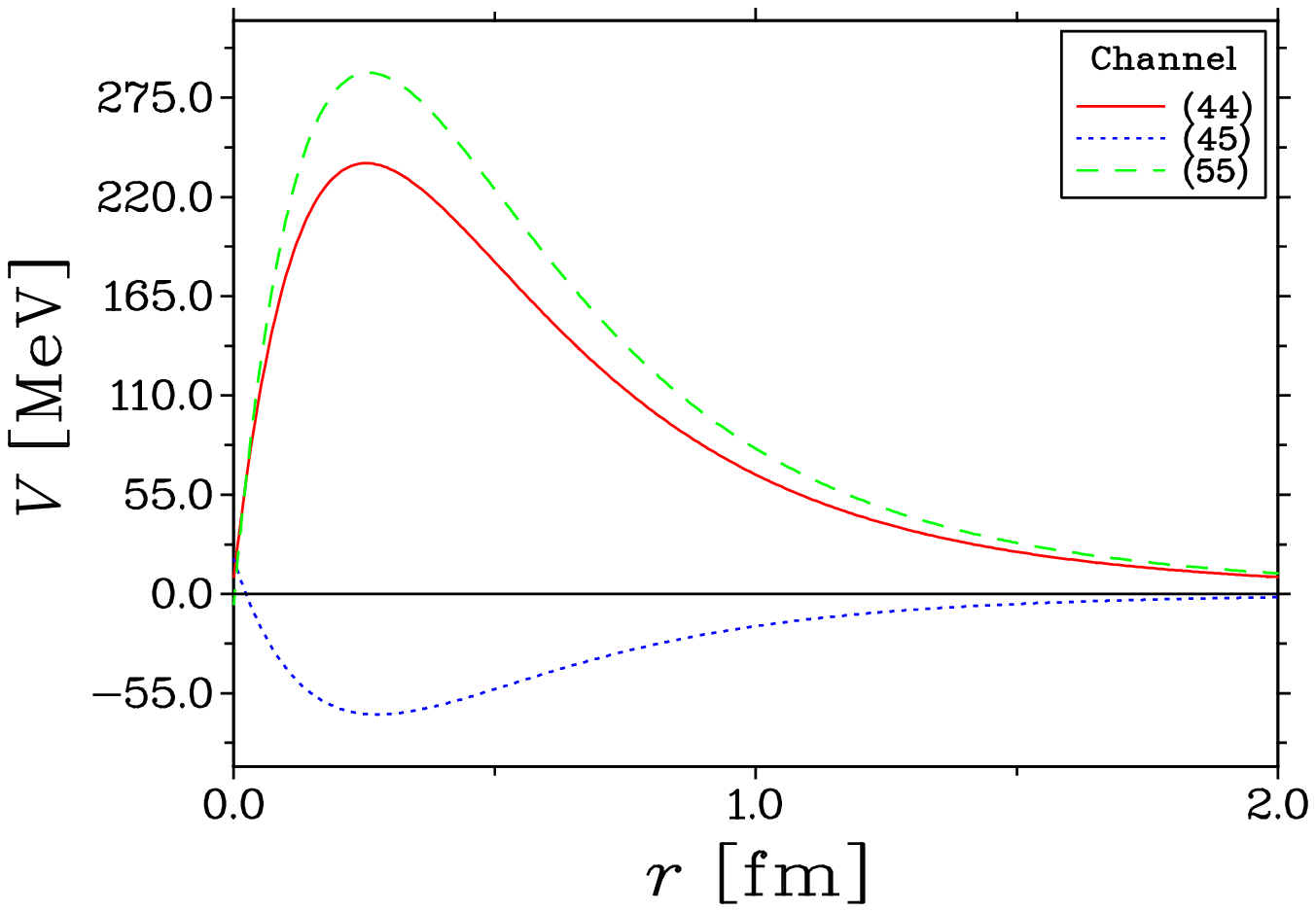}}&\scalebox{0.6}{\includegraphics{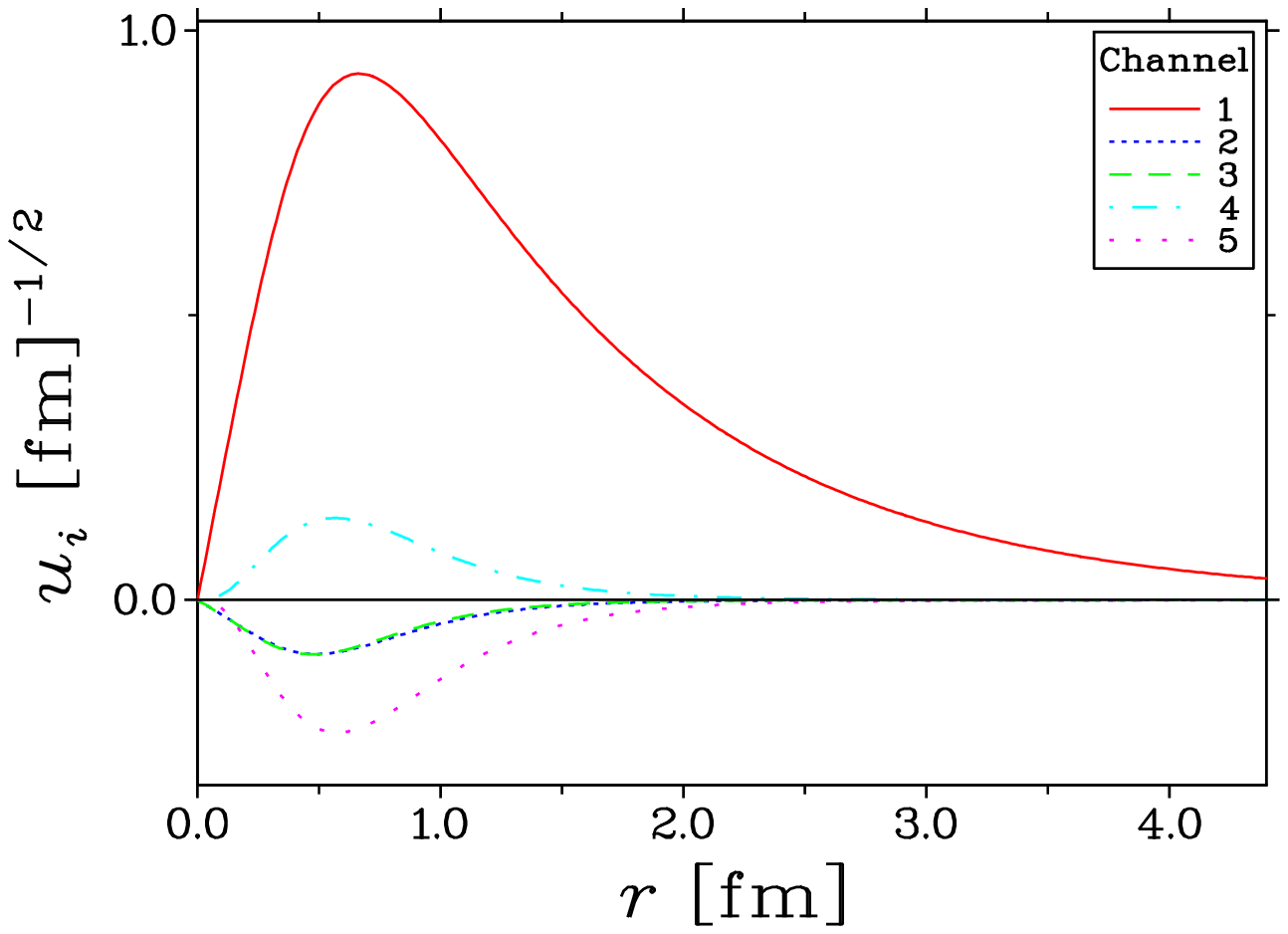}}\\
(c)&(d)
\end{tabular}
\caption{Diagrams (a), (b), and (c) show diagonal and transition potentials of $S$-wave case, $S$-$D$ transition case, and $D$-wave case with the cutoff parameter $\Lambda_\pi=1.0$ GeV, respectively. $(ij)$ denotes the potential $V_{ij}(\Lambda_\pi,r)$. The last diagram shows wave functions with the cutoff parameter $\Lambda_\pi=1.1$ GeV.}\label{plotVW}
\end{figure}

For the hadron masses, we use $m_\pi=137.27$ MeV, $m_{\Lambda_c}=2286.46$ MeV, $m_{\Sigma_c}=2453.56$ MeV, and $m_{\Sigma_c}^*=2517.97$ MeV \cite{PDG2010}. In Fig. \ref{plotVW} (a), (b), and (c), we plot diagonal and transition potentials of $S$-wave case, $S$-$D$ transition case, and $D$-wave case with the cutoff parameter $\Lambda_\pi=1.0$ GeV, respectively. From the diagrams, it is obvious that the tensor forces are strong and thus the coupled channel effects may be important. Another observation is that all the diagonal potentials are repulsive in this simple model. Therefore, the binding solution would result purely from coupled channel effects.

\begin{table}[htb]
\begin{tabular}{c|cccccc}\hline
$\Lambda_\pi$ (GeV)&        1.0 &  1.1 &1.2&1.3&1.4&1.5\\\hline
B.E. (MeV)&3.39& 14.45&35.44& 68.37& 115.06&177.07\\
$\sqrt{\langle r^2\rangle}$ (fm)& 2.0 &1.2&0.9&0.7&0.6&0.5\\
Prob. (\%)&(97.4/0.2/0.2/&(94.3/0.5/0.5/& (90.7/1.1/1.0/ & (86.8/1.8/1.8/ & (82.8/2.6/2.8/ & (79.0/3.4/3.9/\\
          & 0.6/1.6)      &1.3/3.4)     &  2.0/5.2)      &  2.6/7.0)      & 3.3/8.5) & 3.8/9.9)\\
$D$-wave prob.& 2.2\% & 4.7\% & 7.2\% & 9.6\% & 11.8\% & 13.7\%\\\hline
\end{tabular}
\caption{Binding solutions for the coupled $\Lambda_c\Lambda_c$ system with 5-channel contributions. Binding energy (B.E.) is given with relative to $\Lambda_c\Lambda_c$ threshold. The probabilities correspond to $\Lambda_c\Lambda_c(^1S_0)$, $\Sigma_c\Sigma_c(^1S_0)$, $\Sigma_c^*\Sigma_c^*(^1S_0)$, $\Sigma_c^*\Sigma_c^*(^5D_0)$, and $\Sigma_c\Sigma_c^*(^5D_0)$, respectively.}\label{fullcal}
\end{table}

\begin{table}[htb]
\begin{tabular}{c|cccccc}\hline
$\Lambda_\pi$ (GeV)&        1.0 &  1.1 &1.2&1.3&1.4&1.5\\\hline
B.E. (MeV)&0.07& 3.86&15.06& 35.90& 68.39& 114.25\\
$\sqrt{\langle r^2\rangle}$ (fm)& 11.3 &1.9&1.1&0.8&0.7&0.6\\
Prob. (\%)&(99.6/0.0/0.0/&(97.0/0.5/0.1/& (93.4/1.5/0.2/ & (89.0/3.1/0.4/ & (84.2/5.2/0.6/& (79.3/7.6/0.9\\
          & 0.4)         &2.4)          &  4.9)          &  7.5)          & 10.0) & 12.2)\\\hline
\end{tabular}
\caption{Binding solutions for the coupled $\Lambda_c\Lambda_c$ system with only one $D$-wave channel. Binding energy (B.E.) is given with relative to $\Lambda_c\Lambda_c$ threshold. The probabilities correspond to $\Lambda_c\Lambda_c(^1S_0)$, $\Sigma_c\Sigma_c(^1S_0)$, $\Sigma_c^*\Sigma_c^*(^1S_0)$, and $\Sigma_c\Sigma_c^*(^5D_0)$, respectively.}\label{onlyD5}
\end{table}

It is not difficult to solve the coupled channel equations using the variational method \cite{Hiyama2003}. We obtain the numerical results in Table \ref{fullcal} with all the five channel contributions. If we drop the $D$-wave channels, we do not find any binding solution. To see the importance of the tensor force, we include the contribution of only one $D$-wave channel. For the case of $\Sigma_c^*\Sigma_c^*(^5D_0)$, one does not find binding solutions for $\Lambda_\pi<1.3$ MeV. The results for the case of $\Sigma_c\Sigma_c^*(^5D_0)$ (but without $\Sigma_c^*\Sigma_c^*(^5D_0)$) are given in Table \ref{onlyD5}. It is clear that the tensor force from $S$-$D$ wave mixing is essential in getting binding solutions. From these results, one concludes that the channel $\Sigma_c\Sigma_c^*(^5D_0)$ plays a more important role than $\Sigma_c^*\Sigma_c^*(^5D_0)$.

\begin{figure}[htb]
\centering
\begin{tabular}{cc}
\scalebox{0.5}{\includegraphics{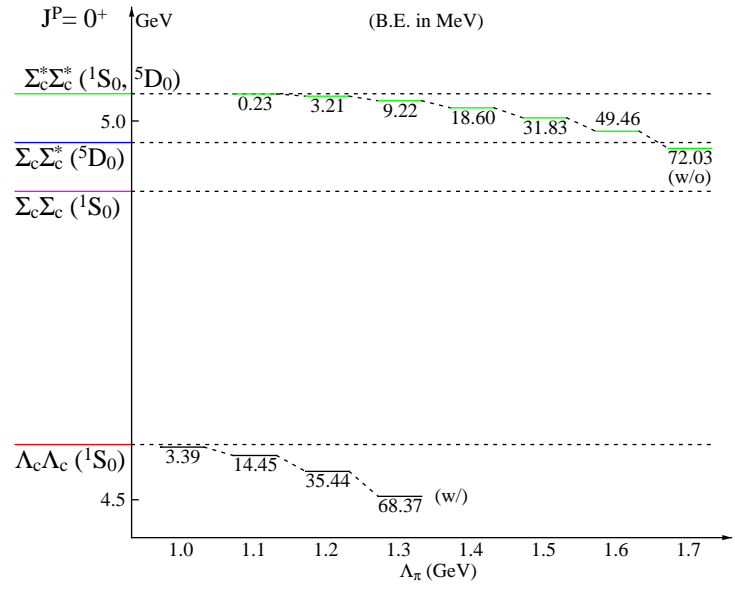}}
\end{tabular}
\caption{Sensitivity of the binding energy on the cutoff parameter $\Lambda_\pi$.}\label{senB}
\end{figure}

As an example of the five channel solutions, we show the wave functions with $\Lambda_\pi=1.1$ GeV in Fig. \ref{plotVW} (d). To see the sensitivity of the binding energy on the cutoff $\Lambda_\pi$, we present a diagrammatic form for the results from Table \ref{fullcal} in Fig. \ref{senB}, where we also show the binding energies for the uncoupled channel $\Sigma_c^*\Sigma_c^*$ (with $S$-$D$ mixing but without mixing $\Sigma_c^*\Sigma_c$ or $\Lambda_c\Lambda_c$ channels). There is no binding solution in other uncoupled channels since the potentials are all repulsive. From Fig. \ref{senB}, if $\Lambda_\pi\geq1.0$ GeV is reasonable, a bound state is possible although the diagonal potentials are all repulsive and there are no binding solutions in individual channels. This indicates the importance of the tensor force.

Although we do not have enough information to determine the cutoff parameter for the heavy quark baryons, we have interesting results in a reasonable range of $\Lambda_\pi$. From the experience of nuclear force, the cutoff should be around 1.0 GeV or larger, depending on the model. For the heavier hadrons, the extended structure is smaller and the cutoff parameter should be accordingly larger. In this study, the solutions corresponding to $\Lambda_\pi=1.0\sim1.2$ GeV are molecule-like because the bound state is not so deep and the inter-hadron distance is not so small. A larger cutoff results in a tightly bound state and the OPEP model may be inapplicable any more. In all, it is possible to have a bound state of two $\Lambda_c$'s while the binding energy is not determined precisely with the present approach. We hope that future studies may specify the binding energy of such a molecule state. On the experimental side, finding the double-charm $\Lambda_c\Lambda_c$ bound state will be a challenging subject at GSI, J-PARC, RHIC, or Belle.

% One should consider intermediate-range scalar meson, short-range vector mesons (or equivalent multi-pion exchange), and the contribution of the $\Xi_{cc}N$ channel with higher cutoff.

%bottom case

In short summary, we have investigated the $S$-wave $\Lambda_c\Lambda_c$ molecule problem by including the coupled channel effects caused by $\Sigma_c$ and $\Sigma_c^*$ in a one-pion-exchange potential model. The couplings to the $D$-wave channels $\Sigma_c\Sigma_c^*$ and $\Sigma_c^*\Sigma_c^*$ are crucial in binding two $\Lambda_c$'s. The results are sensitive to the cutoff parameter $\Lambda_\pi$. If the model cutoff around $\Lambda_\pi=1.0\sim1.2$ GeV is reasonable, one gets a molecule-like solution.

Note: In a recent paper \cite{Lee2011}, the authors also studied $\Lambda_c\Lambda_c$ system but they did not consider the excited $\Sigma_c^*$ contributions. The omission of $D$-wave channels results in different conclusions.

%%%%%%%%%%%%%%%%%%%%%%%%%%%%%%%%
\section*{Acknowledgments}
%%%%%%%%%%%%%%%%%%%%%%%%%%%%%%%%

This project was supported by the
Japan Society for the Promotion of Science under Contract No.
P09027; KAKENHI under Contract Nos. 19540275, 20540281, 22105503, and 21$\cdot$09027.


\begin{thebibliography}{99}

%
\bibitem{Liu2011}Yan-Rui Liu, Makoto Oka, arXiv: 1103.4624 [hep-ph].
\bibitem{Riska2005}F. Fr\"omel, B.Juli\'a-D\'iaz, D. O. Riska, Nucl. Phys. A {\bf 750}, 337 (2005); B.Juli\'a-D\'iaz, D. O. Riska, Nucl. Phys. A {\bf 755}, 431c (2005).
\bibitem{Yan1992}T.M. Yan et al., Phys. Rev. D {\bf 46}, 1148 (1992), ibid {\bf 55}, 5851 (1997).

\bibitem{PDG2010}K. Nakamura et al. (Particle Data Group), J. Phys. G 37, 075021 (2010).

\bibitem{Bonn}R. Machleidt, K. Holinde and Ch. Elster, Phys. Rep. {\bf 149}, 1 (1987).
\bibitem{Julich}B. Holzenkamp, K. Holinde and J. Speth, Nucl. Phys. {\bf A500}, 485 (1989); A. Reuber, K. Holinde, J. Speth, Nucl. Phys. {\bf A570}, 543 (1994).
\bibitem{Hiyama2003}E. Hiyama, Y. Kino and M. Kamimura, Prog. Part. Nucl. Phys. {\bf 51}, 223 (2003).

\bibitem{Lee2011}Ning Lee, Zhi-Gang Luo, Xiao-Lin Chen, and Shi-Lin Zhu, arXiv: 1104.4257 [hep-ph].

\end{thebibliography}
\end{document}